\RequirePackage[hyphens,spaces,obeyspaces]{url}
\documentclass[sigconf]{acmart}
\usepackage{epsfig}
\usepackage{subcaption}
\usepackage{calc}

\usepackage{multicol}
\usepackage{pslatex}
\usepackage{apalike}
\usepackage[ruled]{algorithm2e}
\usepackage[bottom]{footmisc}

\usepackage{longtable}

\usepackage{color}
\usepackage{colortbl}
\definecolor{Gray}{gray}{0.9}

\usepackage{graphicx}
\usepackage{framed}
\usepackage{tabularx}
\usepackage{enumerate}
\usepackage{endnotes}
\usepackage{etoolbox}
\patchcmd{\enoteformat}{1.8em}{0pt}{}{}

\let\footnote=\endnote

\usepackage{comment}
\usepackage{pifont}
\usepackage{acronym}
\usepackage{color,colortbl}
\definecolor{Gray}{gray}{0.9}
\definecolor{DarkGray}{RGB}{96,96,96}

\usepackage{mdframed}
\usepackage{subcaption}
\usepackage{fixltx2e}
\usepackage{stfloats}
\usepackage{amsmath}%
\usepackage{balance}
\usepackage{hyperref}
\usepackage[title]{appendix}
\usepackage{color,soul}
\usepackage{textcomp}
\usepackage{manyfoot}
\usepackage{booktabs}
\usepackage[english=american,threshold=40,thresholdtype=words,csdisplay,autopunct]{csquotes}

\setcopyright{none}
  {\list{}{\leftmargin=0.05in\rightmargin=0.05in}\item[]}%
 {\endlist}

\usepackage{array}
   \newcolumntype{C}[1]{>{\centering\arraybackslash}m{#1}}
      \newcolumntype{L}[1]{>{\raggedright\arraybackslash}m{#1}}

\def\lq{``}
\def\rq{''}
\newcommand\qqq[1]{\lq{}{\emph{#1}}\rq{}}

\copyrightyear{2025}
\acmYear{2025}
\setcopyright{rightsretained}
\acmConference[NSPW '25]{New Security Paradigms Workshop}{ 2025}{Location}
\acmBooktitle{New Security Paradigms Workshop (NSPW '25), Location}
\acmDOI{DOI}
\acmISBN{ISBN}

\acrodef{SoK}{Systematization of Knowledge}
\acrodef{HCI}{Human Computer Interaction}
\acrodef{DCS}{Developer-Centered Security}
\acrodef{PETs}{Privacy Enhancing Technologies}

\begin{document}

\title{\textbf{\emph{Extended Version of Paper Presented at ICISSP, Porto 20-22 February, 2025}}\\
\ \\
A Value-Driven Approach to the Online Consent Conundrum - A Study with the Unemployed}
%\author{Paul van Schaik, Karen Renaud\\
%Teesside University, University of Strathclyde\\
%p.van-schaik@tees.ac.uk, karen.renaud@strath.ac.uk}
\date{}

\author{Paul van Schaik}
\orcid{https://orcid.org/0000-0001-5322-6554}
\affiliation{%
  \institution{Teesside University}
  \streetaddress{}
  \city{Middlesbrough}
 % \state{}
  \country{UK\\}
 % \postcode{}
}\email{p.van-schaik@tees.ac.uk}

\author{Karen V. Renaud}

\orcid{https://orcid.org/0000-0002-7187-6531}
\affiliation{%
  \institution{University of Strathclyde,}
  %\streetaddress{}
  \city{Glasgow}
  %\state{}
  \country{UK\\}
  %\postcode{}
  \institution{University of South Africa, RSA,  Rhodes University, RSA}
}\email{karen.renaud@strath.ac.uk}

\begin{abstract}
Online services are required to gain informed consent from users to collect, store and analyse their personal data, both intentionally divulged and derived during their use of the service. 
 There are many issues with these forms: they are too long, too  complex and demand the user's attention too frequently. Many users consent without reading so do not know what they are agreeing to. As such,
 granted consent is effectively uninformed. In this paper, we report on two studies we carried out to arrive at a value-driven approach to inform efforts to reduce the length of  consent forms. The first study  interviewed unemployed users to identify the values they want these forms to satisfy. The second survey study helped us to quantify the values and value creators. To ensure that we understood the particular valuation of the unemployed, we compared their responses to those of an employed demographic and observed no significant  differences between their prioritisation on any of the values. However, we did find substantial differences between values and value creators, with effort minimisation being most valued by our participants.
\end{abstract}

%\keywords{Values, Value Creators, Consent Forms, Unemployed} %MANDATORY
 % \noindent Target journal: Journal of Privacy and Confidentiality

%    The main manuscript should be as succinct and readable as possible. 20 pages is a good length.
%    The supplementary material can be any length (and in some cases may be unnecessary). The supplementary material may contain additional figures and tables, details about data, mathematical derivations, further explanations, or additional simulations or other supporting studies. The supplementary material should be formatted using the same stylesheet, and uploaded separately.

\maketitle

%\textcolor{red}{I have difficulty with the unemployed vs employed. We did not ask the employed for their values, but asked them to rank the values the unemployed gave us. Is this acceptable? It might cause criticism and I am not sure how to resolve this issue. We just seem to drop the employed group in for comparison without justifying or arguing for it. }

%\textcolor{red}{We mention comparing employed and unemployed and, in the results, explain that in Study 2 there was no significant effect or interaction effect  of employment status on any of the  measures}

\section{Introduction } %KAREN
The requirement for mandating informed consent to permit online data gathering and processing  inherits the paradigm from the fields of medicine and research \citep{beauchamp2011informed}. Legislation such as the European Union's General Data Protection Regulation (GDPR) forces online service providers to ask users to consent to collection, storage and processing of their data. There are, unfortunately, many reasons for the failure of this mechanism to obtain truly informed consent. 

 \citet[p. 1888]{solove2012introduction} writes that “consent is not meaningful in many contexts involving privacy” because \qqq{(1) people do not read privacy policies; (2) if people read them, they do not understand them; (3) if people read and understand them, they often lack enough background knowledge to make an informed choice; and (4) if people read them, understand them, and can make an informed choice, their choice might be skewed by various decision making difficulties}.
Solove is suggesting that people are not granting \emph{informed} consent.
Users cope with the frequent  unusable consent forms they encounter by dismissing them (without reading them) \citep{parfenova2024words}. This means that consent forms, in general,  do not fulfil their core purpose \citep{chomanski2023online}. 

The previous paragraph describes the current status quo in terms of consent forms (related research is summarised in Section \ref{BG}). The
length of these policies deters online users from engaging with them (\cite{mcdonald2008cost}). If the length can be reduced, it  would likely mitigate the situation, but such shortening must  be done mindfully. One way to do so is to ensure that the forms provide only the information that satisfies users' values. As yet, we do not know what these values are, and in the absence of this, consent forms strive towards comprehensive coverage of all information. 

We  identified these values  by interviewing unemployed users. Having derived a set of values from a qualitative analysis, we carried out a second study to determine how users (specifically our target users, the unemployed) would comparatively rate the derived values (Section \ref{methodology}). 
(Section \ref{Findings} reports on our findings.
These findings make it possible to reduce the length of consent forms in such a way that the provided information aligns with users' values. Anything else  can be  made available on demand. 

\begin{figure}[h]
    \centering
    \includegraphics[width=\linewidth]{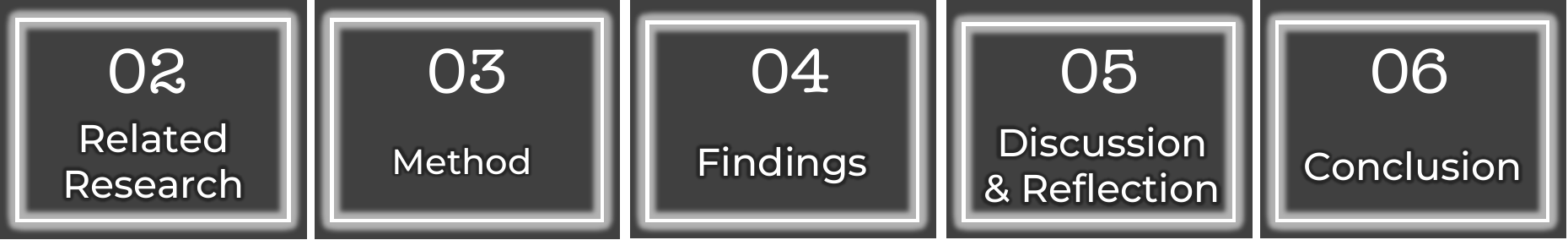}
    \caption{Structure}
    \label{fig:flow}
\end{figure}

Figure \ref{fig:flow} depicts the structure of this paper.

\section{Related Research} \label{BG}%KAREN
To provide this review of related research, we carried out a
search as follows (Search tallies are presented in  \ref{tab:searches} witin the Appendix.):\\
 \emph{Databases:} SCOPUS, ProQuest, ACM DL, EBSCO, SSRN, Ingenta, DBLP, Mendeley, SpringerLink;\\
 \emph{Years:} 2013 onwards;\\
 \emph{Keywords:}
“online consent” + privacy;\\
 \emph{Exclusions:}  Cookie consent, online health, 
 research participation;\\
 \emph{Inclusions:} Peer reviewed, English, Dutch or German.
\\
\ \\
\noindent\textbf{Summary of Literature}\\
 \citet[p. 2]{obar2022older} refers to the impossible  position the online user is in, being  \qqq{both perpetual data subject and overwhelmed privacy savior}. \citet{fassl2021stop} calls the current situation `Consent Theater'.
\citet{richards2018pathologies} wonder if the current consent granting paradigm is `too much of a good thing'.
 \citet{zong2020individual} agrees, arguing that what passes for consent does not provide the protections  it purports to provide. There is a clear need to       minimise `unwitting' consent \citep{richards2018pathologies}.

%Human \etal \cite{human2022data} enumerates a range of challenges of Data Protection and Consenting Communication Mechanisms:

The papers resulting from the search were sorted into five categories. The first three  mirror the \citet{solove2012introduction} conceptualisation of the issues with consent forms.
%{\color{red}*****Remove misleading 3D representation.*****}
%
%\begin{figure}[h]
%    \centering
%    \includegraphics[width=\columnwidth]{figs/LIT.pdf}
%    \caption{Overview of Literature Topics}
%    \label{fig:LIT}
%\end{figure}

% \noindent\textbf{(1) Power imbalance in motivation for  the consent form (prioritises service provider instead of user):}\\
% To re-establish power balance, forms should:
% \begin{itemize}%

%\end{itemize}
\ \\
 \noindent\textbf{(1) Difficulties READING the consent form:}\\
 \citet{bustos2014we} find that consent forms are unread by the vast majority of online users. Issues contributing to this include:
%{\color{red}*****Problem: inconsistency (1) is phrased in terms of barriers, but (2) and (3) are phrased in terms of solutions.  Readers demand and deserve better than this inconsistency.*****}

\begin{itemize}
    \item  Forms do not display concise, comprehensible, unambiguous, complete information   \citep{human2022data,carvalho2018express,carolan2016continuing,smith2016transforming,kani2012service}.
    \item Forms are not always upfront with about how customer data are collected and shared \citep{desai2003commerce}.
      \item Forms lead to a sense of information overload  \citep{human2022data}.

\end{itemize}
\ \\
 \noindent\textbf{(2) Difficulties UNDERSTANDING the consent form:}\\
Even if users read the forms, they do not always understand them:
\begin{itemize}
\item The language might exceed what a functionally literate adult can be expected to understand \citep{luger2013consent}. \item Complex language used to communicate information  \citep{human2022data} 
\item The required skills and knowledge to process consent form contents are not possessed by  users \citep{van2012online}, which includes overly technical terminology and complex privacy concepts \citep{rowan2021comprehension}.
 \item No just-in-time notifications to ensure that people get the information as and when they need it \citep{shulman2023informing}.
\end{itemize}
\ \\
\noindent\textbf{(3) Difficulties INTERPRETING the information provided by the forms:}\\
Users might not get all the knowledge they need. Missing information might be:
\begin{itemize}
\item No mention of accountability, auditability, transparency and repudiation \citep{carvalho2018express,nguyen2022freely,human2022data}.
    \item No commitment to respecting privacy \citep{smith2016transforming}.
    \item Not communicating privacy risks with a configurable dialogue \citep{kani2012service,fox2022enhancing} and not providing a projection of future consequences \citep{rowan2021comprehension}.
 \item Not explaining how to obtain post-consent access to information and decisions  \citep{human2022data,carvalho2018express} nor mention of the right to withdraw consent \citep{carvalho2018express,richards2018pathologies}. %This means that users would have to provide  proof of identity  when consenting \citep{human2022data}.
 \item No social annotation functionality provided\citep{balestra2016effect}.
\end{itemize}

\ \\
\noindent\textbf{(4) Challenges:}
\begin{itemize}
\item Coerced or incapacitated consent \citep{richards2018pathologies,carvalho2018express} leads to  imbalance of power  \citep{human2022data}.
\item Knowledge asymmetry exists between users and online service providers \citep{bashir2015online}.
\item Digital service providers display long and complicated privacy policies that do not engage individuals nor help them understand the implications of agreement \citep{obar2022older}.
\item There is potential for exploitation, which mandates transparency 
\citep{hutchinson2022challenges}.
      \item Users are not aware, but should be made aware, of the influences of nudges and nudge positioning on privacy choices \citep{utz2019informed}.   Nudging children towards privacy-invasive choices might be unethical if it triggers oversharing  \citep{veretilnykova2021nudging}.

\item Balancing individuals' privacy rights against commercial access to personal information needs to be resolved \citep{mohamed2012not}.
\item Positive emotions might lead to disregarding of important information \citep{shulman2023informing}.

 \item  Users' cognitive limitations are not, but should be accommodated \citep{human2022data} 
 \item  Best practice is not, but should be enforced \citep{human2022data}, for example data minimisation \citep{carvalho2018express}.
 \begin{itemize}
    \item Users' preferences containing personal data  \citep{human2022data}
 \item  Legal requirements  \citep{human2022data}
 \item  Standardisation  \citep{human2022data}
 \item Confirmation  of assent to be sent by email \citep{carvalho2018express}.
\end{itemize}
\end{itemize}

\ \\
\noindent\textbf{(5) Solutions:}
\begin{itemize}
\item \citet{belanger2013pocket} propose a mechanism whereby parents can consent on behalf of their children under the age of 13 before online entities collect information from them. 
\item Online informed consent reading can be improved with an AI-Powered Chatbot \citep{xiao2023inform}.
    \item \citet{burkhardt2022privacy} suggests a model of informed consent grounded in  the theory of planned behaviour (TPB), which maximises autonomy and the need to be fully informed.
    \item \citet{jarovsky2018improving} proposes MAPP
(Methodology for Autonomy-Preserving Protection), a methodology to evaluate  designs that are intended  
 to improve consent  decisions.
 \item \citet{bustos2014we} suggests that online consent be replaced by a set of applied cryptography tools rather than trying to improve the consent process. 
 \item {Collective Refusal} can help people address the autonomy problem, and utilise the collective power of all users \citep{zong2020individual}.
 \item \cite{rooney2018online} suggest an `ethics of virtue' approach to achieve informed consent.
\end{itemize}

%\begin{itemize}
 %   \item 
%Human-centric and Human Computer Interaction
%\begin{itemize}
% \item  Imbalance of power  \citep{human2022data} i.e., must be freely given \citep{carvalho2018express}.
% \item  Respect of User Constraints  \citep{human2022data}
 %\item  Display concise, comprehensible, unambiguous, complete information   \citep{human2022data,carvalho2018express,carolan2016continuing}.
% \item  Enforce Good Practices  \citep{human2022data} such as data minimisation \citep{carvalho2018express}.
%\end{itemize}
%\item Accountability, Auditability and Transparency \citep{carvalho2018express}.

%\item Legal

%\item Technical
%\begin{itemize}
%    \item Technological variety  \citep{human2022data}

% \item  Communication of information  \citep{human2022data}
%\end{itemize}
%\end{itemize}
\ \\
\ \\
\noindent\textbf{Motivation for Value-Driven Approach}
This short review paints a complicated picture, with many issues and challenges related to the current consent-granting process. 
\citet{richards2018pathologies} argue for three circumstances to be necessary for an ideal environment for effective consent. They argue that consent requests: (1) should be infrequent, (2) the risks of giving consent must be vivid and easy to envision, and (3) data subjects must have an incentive to take each request seriously. 

With respect to \citet{richards2018pathologies}'s  first point, it might be argued that legal change is needed to reduce the frequency of consent requests. The second aspect might be possible to address with an intervention to ensure that the risks are clearly communicated. The third point is moot, as people have been lulled into divulging a great deal of their information and the inertia effect might well make it difficult to turn that situation around to incentivise them  to process consent forms thoughtfully.

The second point is arguably linked to the increasing length of these consent forms.
This might be because the usual approach to improving online consent forms is to provide \emph{more} information, thereby increasing length and exacerbating verbosity \citep{klotz2021subtract,haley2022illusory,stamey2009automatically}. 
The underlying implicit
assumptions by those who craft online consent forms that emerge are that: (1) people want \textbf{all} possible information about how their personal data will be
stored and used, (2) decision-making can only be improved if exhaustive information is provided,
and (3) liability must be limited by having trained legal staff craft online consent forms. However, these assumptions are
unfounded \citep{buchanan2001information,mackinnon1980complexity,gigerenzer2017cassandra}.

 However, \citet{sunstein2020too} argues that people are easily overwhelmed by
too much information, which explains why they struggle with  increasingly lengthy online consent forms.
%Moreover, the
%involvement of legal entities in crafting online consent forms has maximised complexity \citep{purcaru2014informed}. 
%
%
%
Instead of `adding' more information,
 \citet{klotz2021subtract} advocates a \emph{subtraction} approach -- paring  down instead of exacerbating complexity. 
 If we seek to subtract, as advocated by  \citet{klotz2021subtract},  we need to gain  insight into what should be retained and what can feasibly be removed. 
 \citet{sunstein2020too} suggests that only information that improves well-being should be included. %This, too, 
%means that information should only be provided if it maximises well-being by satisfying people’s
%needs. 

In suggesting that a subtraction approach might be viable, we align with 
  \citet{guthrie1996religion}, who argues that people often do not need or want more information;  they want  \emph{the right kind of information}. %If we cut
%down ICDs, how do we decide what to retain? 
Legal requirements have to be
satisfied, while ensuring that well-being is maximised according to Sunstein.
%A viable approach would distil the \emph{essence} of the information in the ICD  \cite{klotz2021subtract}. 

%To apply this approach, we need first to
%understand the priorities of online users. That will allow us to provide prioritised information in a
%comprehensible way and then provide further detail on request via clickable link URLs \cite{renaud2018make}.  

Our proposal is to design online consent forms in such a way that they  satisfy the needs of online users. To that end, 
we need first to find out what people’s actual values are.
This is important because providing information that aligns with their values is likely the best way to maximise well being. 

%As such, the idea we explore in this paper is to applying a  address the second of Richard and Harzog's points by applying a`\textbf{Subtract and provide info that builds well-being}' approach. 

It is important to establish both \emph{values} and \emph{value creators}, as the relevance or priority of each might well differ between domains. We apply the laddering  for identifying values \citep{moghimi2016iranian} and value creators and the analytic hierarchy process (AHP) for quantifying the relative importance of values and value creators \citep{saaty1987analytic}.  Laddering interviews are  used to identify a qualitative hierarchy of values and value creators. For example, research in housing has used means-end chain analysis to study and identify human needs, conceptualised as values, as well as value creators that contribute to fulfilling these needs, as a basis for designing and evaluating homes from the perspective of their inhabitants \citep{moghimi2017incorporating}. The AHP quantifies the hierarchical relations by way of rated pairwise comparisons. For example, the AHP was used to quantify the relative importance of patients’ preferences and priorities regarding colorectal cancer screening \citep{dolan2013patients}.

Consider that  the design and evaluation of online-games, social housing or online-consent all have distinct features and satisfy different needs. Our focus is on values that apply to the online consent context. \\
\ \\
\noindent\textbf{Studied Demographic}\\
 We chose to focus on values of the unemployed, and their experiences in engaging with online consent forms.
 \citet{seabright2010company} explains that the unemployed inhabit `information islands'. Unlike the employed who benefit from regular security awareness training, there are no bridges for the unemployed to gain up-to-date information. This means that it is easy for misunderstandings to gain traction because people are out of touch with the latest security advice. Seabright says that society does not construct bridges to  increasingly isolated unemployed communities. The cyber security field is dynamic and fluid due to the sustained and inventive efforts of  cyber criminals. This demographic is thus more vulnerable to losing their privacy.
Moreover, declining to give consent might be infeasible if monetary rewards are dependent on consent, perhaps more likely a pressure point for the unemployed. \\
\ \\
\noindent\textbf{Research Questions}\\
\noindent\textbf{Study 1:} What are the informed consent-related values and value creators for the unemployed?

\noindent\textbf{Study 2: }
%\indent (a) How do \textit{unemployed} people weight the values and value creators in the informed consent context?
%\\
\noindent How do {unemployed} people weight the values and value creators in the informed consent context, and how do these differ from the way the employed rank these?

\section{Method } \label{methodology}%PAUL
%\subsection{Research Design}
\emph{Study 1.} A laddering interview design \citep{dolan1989medical} was used to elicit values and value creators that contribute to informed consent in the context of online services. The \emph{output} of the study was a hierarchical set of values and value creators. We also measured privacy literacy \citep{Trepte2015}.

\emph{Study 2.} A two-group independent measures design was used. The groups were unemployed and employed people. The \emph{output} of the study was a quantified hierarchical structure of values and value creators with value and value-creator weightings separate for employed vs unemployed participants.
%In addition, we measured privacy literacy and affinity for technology interaction.
\\
\ \\
\noindent\textbf{Participants}\\
\emph{Study 1.} Thirteen unemployed participants responded to an invitation from a previous study (\cite{van2024privacy}). They were compensated for their time through a voucher or a SIM card twith gigabytes of free data to use on their smartphone. 
\par\emph{Study 2}.  One hundred and two unemployed participants and the same number of employed participants, 115  female (68 unemployed, 47 employed) and 89 male (34 unemployed, 55 employed), were recruited through via an online survey panel and compensated for their time according to the panel's rate.  Their mean age was 52 (\emph{SD} = 15). 
\\
\ \\
\noindent\textbf{Materials}\\
\emph{Study 1}. A laddering interview guide was created and used.  Two scales were used: the Online Privacy Literacy Scale (OPLIS) \citep{Trepte2015} was used to measure specific privacy knowledge. (Details of our  OPLIS scoring scheme are presented under Study 2, as the development of this scheme required a larger sample than that of Study 1.) In this sample the mean was 12.92 (\textit{SD} =3.12), which corresponds to approximately a 57 percentile rank according to \cite{masur2017entwicklung}.
\par\emph{Study 2}. An AHP survey was created according to guidelines for survey construction \citep{dolan1989analytic,Dolan2008418}, using the output of Study 1 (a hierarchy of values and value creators) as input.  All the possible pairs of values underlying the higher-order goal of informed consent were formed as well as all the possible pairs of value creators underlying each value.  The collective pairs were presented sequentially (first the value pairs [randomised] and then the value creator pairs for each value [values randomised; value creator pairs randomised within values]). Participants had to evaluate the relative importance within each pair (for example, the importance of the value of control relative to that of fairness). As in Study 1, OPLIS measured privacy literacy. The Affinity for Technology Interaction scale (ATI) measured \qqq{the tendency to actively engage in intensive technology interaction} \citep[p. 456]{Franke2019456}.
\\
\ \\
\noindent\textbf{Procedure}\\
\emph{Study 1.} Because of pandemic restrictions and to reach a geographically UK-wide audience, interviews were conducted remotely by VoIP or by telephone, recorded and automatically transcribed. Afterwards, the recordings were played back and any corrections were made to the transcripts. 
In each interview, the participant was asked to identify value creators (\emph{what} an online consent form should provide) and subsequently for each value creator one or more values (\emph{why} the value creator is  important). Interviews lasted from 14 to 34 minutes. After each interview, the participant was directed to an online survey to complete OPLIS.

\par\emph{Study 2.}  Participants were directed to an online survey that administered demographic questions, a series of AHP pairwise comparisons, OPLIS and the ATI scale.  
\par\emph{Ethics.} Research ethics approval was obtained from the University of Strathclyde and from REPHRAIN, National Research Centre on Privacy, Harm Reduction and Adversarial Influence Online.\\
\ \\
\noindent\textbf{Data analysis}\\
\emph{Study 1.} We used the framework of means-ends chain analysis to identify people's needs (value) and how these could be achieved (value creators) \citep{kilwinger2021methodological}.  Both researchers coded an initial set of five transcripts.  Their individual coding schemes were discussed and a final coding scheme was agreed.  One of the authors then coded all the transcripts.
\par\emph{Study 2.} AHP analysis of response consistency and weightings was conducted \citep{dolan1989analytic}.  Analysis of variance (ANOVA) techniques were used to analyse differences between unemployed and employed participants on the AHP weightings.

\section{Findings}\label{Findings}
%\subsection{OPLIS \& ATI}
The original OPLIS has four dimensions to measure online privacy literacy: institutional practices, technical aspects of data protection, data protection law, and data protection strategies.  As explained by \citet{edelsbrunner2022model},  knowledge in different domains are often best assessed with formative measurement.  Therefore, we created a measure consisting of items from each of these four dimensions.  From each dimension, we selected two items, based on the percentage of the sample that answered correctly:  between 25 pct and 50 pct, the item with the minimum correctness  and,  between 50 pct and 75 pct,  the item with the maximum correctness. This procedure ensured an equal mix of more difficult and easier items. {A \textit{t} test showed that online privacy literacy did not differ between employed (\emph{M} = 56.86; \emph{SD} = 23.57) and unemployed (\emph{M} = 53.19; \emph{SD} = 20.79) participants, \textit{t} = 1.18, \textit{df} = 198.89, \emph{p} = 0.24, d = 0.17.}

Factor analysis of the Study 2 data was conducted on the ATI and produced a one-factor solution, with 56 pct of variance explained. Factor scores were calculated and used in subsequent analysis.  {A \textit{t }test showed that afﬁnity for technology interaction was higher in employed (\emph{M} = 0.14; \emph{SD} = 1.02) than in unemployed (\emph{M} = -0.18; \emph{SD} = 0.91) participants, \textit{t} = 2.84, \textit{df} = 199.31, \emph{p} = 0.005, d = 0.40.}

\subsection{STUDY 1 } \label{study1}
\begin{figure*}[h]
    \centering
    \includegraphics[width=\textwidth]{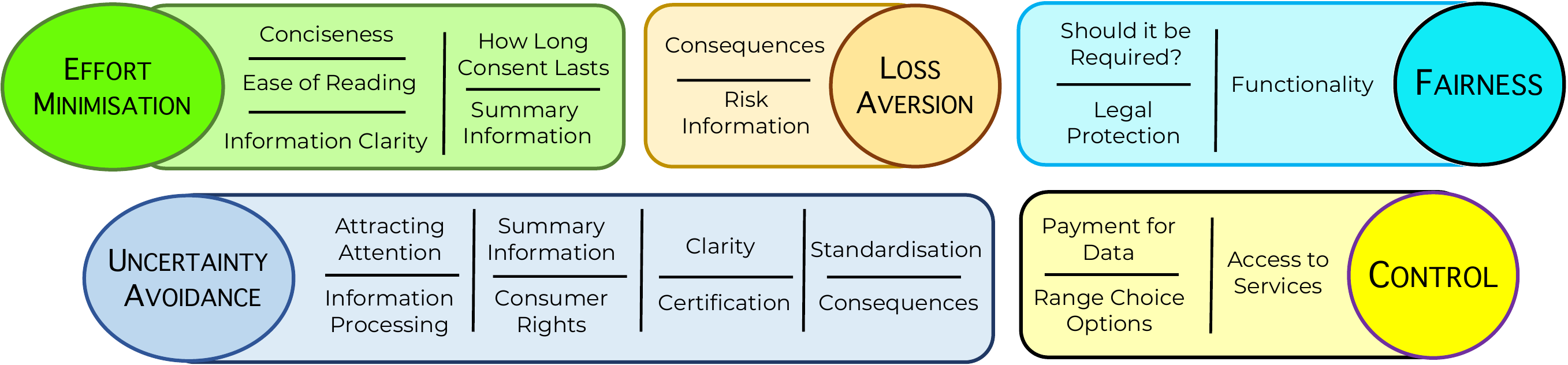}
    \caption{Means-end hierarchy of user values and value creators contributing to informed consent decision making}
    \label{tab:meansEndTable}
\end{figure*}

Study 1 sought to identify a hierarchical means-end structure of informed consent for online services 
%\color{red}*****Table reference does not work properly. 
(Figure \ref{tab:meansEndTable}).

Five values were identified: (1) control, (2) uncertainty avoidance, (3) loss aversion, (4) effort minimisation and (5) fairness.   Under each value, two or more \emph{value creators} were identified.  The value creators (means) underlying each value (end) would contribute to the value.  In turn, the values would contribute to the higher-order goal of informed online consent decision-making.  The values and underlying value creators are presented here, with illustrative quotations, where P{\textless} number{\textgreater} \leavevmode\nobreak\ represents a quoted numbered participant.

\subsubsection{Control}
\emph{Control} was defined as a user's feeling that they have control.  Three value creators contributing to the value of control were identified.\\
\ \\
\emph{\textbf{(1) Payment for Data}} (being paid for giving data/having one’s data captured by an online service) 
 Participants expected to be paid for the data they provided, but this was not often not the case:
 \qqq{I'm a big fan, by the way, of this idea that they pay me for it ... If they're making money off it, I should get my cut.} (P55)
\ \\
\emph{\textbf{(2) Access to services}} (not having to sign up for certain services in order to be able to read the pages) (see \emph{Fairness} [value], \emph{Should consent be required}[value creator])\\
\ \\
\emph{\textbf{(3) Range of Choice Options}} (having a range of choice options in responding to an online consent form, for example, only the options of accept all or reject all, or a larger set of more fine-grained options that offers more choice).

 Participants expected they would be able to select which data will be shared:
 \qqq{What I want to see in these forms is  ... can I selectively choose? Alright, you can have my location, but you can't have something else. You can't track me for example, as I'm using the website.} (P26)
% \qqq{Some of them ask you to either accept everything or reject everything or otherwise go to a different screen where you can select one by one ... that would probably only be necessary if there's a lot of things to choose from.}(P29)

 %They also identified information that had been captured to which they had not consented:
 \qqq{You see particulars, my location or something, but I don't remember consenting to that.} (P26)

%Participants realised that using a service required them to consent to the conditions, even if they might not agree:
% \qqq{Then at the end of it, if you want to use that service you will have to}
%They had noticed a change in consent choice options since the introduction of the Data Protection Act 2018 (the  UK's implementation of the General Data Protection Regulation [GDPR]):
% \qqq{I've noticed a big change in over the last couple of years. Used to just get a yes or a no and sometimes didn't even get a no. But it was just kind of opt in, but now some of them are quite good, you're allowed to opt out or say 'yes' and others when you click opt out you get a list of probably 10 to 20 individual things ...  halfway through you're just looking for the opt-out button and quite honestly, you just click 'yes' most of the time because it's so hard work ... when it changed [to GDPR] and you started to see more of the kind of tick box opt-out.}(P55)

% \color{red} #################################################### inconsistency - DELETE

% \qqq{I don't feel like the cookie thing pops up as often as it used to either ... I don't think they're allowed to do this. I thought that's when it changed, but now you almost feel like you're auto-enrolled onto the website, so ... it's given consent.}(P55) 
% #################################################### \color{black}
% \\

\subsubsection{Uncertainty Avoidance}
\emph{Uncertainty avoidance} was defined as a user's desire to gain information to reduce or remove uncertainty. Eight value creators contributing to the value of uncertainty avoidance were identified.
agree to it and otherwise they would not offer the service.(P29)\\
\ \\
\emph{\textbf{(1) Consumers' Rights}} (information about consumers' rights when they use the specific online service that is provided) 

 Consumers' rights in consent documents were seen as beneficial for  both users and service companies:
 \qqq{I suppose it should provide what what they expect of me and what I can expect of them.} (P56)
 \qqq{I would like to know what my rights are as a consumer so if there was just a sheet of main bullet points that  probably would be better.} (P73)\\
 %\qqq{‘Real’ benefit: will allow people to ascertain their rights when things go wrong. Benefit for company from being honest and straightforward in their online consent forms: customers’ good will/good feeling towards the company/online service provider ... Loyalty: users are more inclined to buy more products and to continue their contract with the company/service provider. Users feeling more protected if things go wrong because they know their rights,  know how to make a complaint or get recompense or where things go wrong.}(P33)
\ \\
\emph{\textbf{(2) Information Processing}} (information about what happens to the user’s personal data)

Participants expected that a consent document would explain how their data would be stored and be processed, and why:
 \qqq{Information in such a consent form  would include how they handle data and privacy [and] whether they hold data on computers in the USA.} (P33)
 \qqq{When information is stored or where it is stored } (P57)
\ \\
\emph{\textbf{(3) Attracting your Attention}} (highlighting important information to attract a your attention)
Participants expected that critical information in a consent document would be highlighted in order to attract users' attention before they decided to consent:
 \qqq{I'm always very of the idea that it should be clear, and if it is very important, it should have something like a red box around it ... should be highlighted ... the ability to follow you to other websites ... The ability to sell the information to unknown and any third party without either compensating me for it or asking my permission. } (P55)
\ \\
\emph{\textbf{(4) Information Clarity}} (how easy it is to understand the online consent text (simplicity and comprehensibility of text) [see effort minimisation].

\emph{\textbf{(5) Consequences}} (information linked to important consequences for the online-service user) [see loss aversion].\\
\ \\
\emph{\textbf{(6) Standardisation}} (standardisation of forms according to relevant information categories for users)

 Standardisation, together with shortening consent documents, would also facilitate users' reading and understanding as a basis for making a consent decision:
 \qqq{If it was shorter; I think if it if it was standardised so you knew what certain sentences meant and so it was just more bullet points and you knew what that referred to instead of it being a long complex thing all the time.} (P77)

\ \\
\emph{\textbf{(7) Certification}} (a sign presented on the consent form to show certification by a trusted party)
 Certification with icon visualization could communicate consent information quickly and clear.
 \qqq{[If] it was icons that you got used to know what they mean, like the Facebook icons. Yes, [if] there was an icon 'we sell your data'. [It] would just be quick when you get used to seeing [the icon], when you know instantly what we were signing up to.} (P77)\\
\ \\
\emph{\textbf{(8) Summary Information}} (summary consent information, with links to further details (for example, clickable icons that link to detailed information) 

 The expectation was that providing summary information with links to detailed information would reduce the required reading time to make a consent decision, make it more likely that users would read the information and cater for users with different needs in terms of information detail:

 \qqq{[A] bit more streamlined so you've got four options, so you don't have to click through. You might have the option to go through and read more information, but you shouldn't have to. It should give you a brief summary of each thing that can select.} (P83)
% \qqq{There's bullet points ... let's say if you want more information, click this and then you'd get presented with all of the information. And then because it just saves a lot of time, I think that's the biggest thing that people don't want to do it for. Well, if I just have a simplified version like this. This is where it's going and how it affects you sort of thing. And I think a lot more people would actually take the time, just doing a quick read through it ... Then it's a lot quicker and a lot easier for them to read it. It's I think that's the underlying thing at the end of the day that stops people from from doing that.}(P57)
% \\ \color{red} ############# this is not privacy - DELETE
% \qqq{I think some people are more detailed than others. I think some people may look at a particular topic like returns. May think: 'Oh I'm not sure whether I want to buy this. I think I'll need to look at this to see what my rights are'. There are people [who] might think that that's not important. I bought (from this company) before and I trust this company. So I think I think you need the main summary and then the opportunity for people to look a little bit more detail if they need to for that particular purchase.}(P73)
%############# \color{black}\\ 

\subsubsection{Effort Minimisation} 
\emph{Effort minimisation} was defined as a reduction in the effort required to process the information that is presented.  Five value creators contributing to the value of  effort minimisation were identified.\\
\ \\
\emph{\textbf{(1) Conciseness}} (conciseness of text, for example, using bullet points) 
  Concise writing in consent documents could facilitate reading and understanding: 
 \qqq{Should follow guidance from the Plain English Campaign. Online consent forms should be brief and concise ... [The] benefit should be easily read and understood in 2 to 3 minutes.} (P33)
\ \\
\emph{\textbf{(2) Ease of Reading}} (how easy it is to read the online consent text, for example in terms of font size) 
 Text should facilitate reading, for example by use of the sufficiently large font, but participants' experience was the opposite:
 %\qqq{large enough to read.}(P56)
 \qqq{Should also make the type (font) bigger. Should use normal print.} (P33)
 %\qqq{They're always in small script. I just cannot be bothered.}(P75)

\emph{\textbf{(3) How long consent lasts}}  (one-time consent process or consent required each time the online service is accessed) 
 Participants experienced the same consent process on repeat use of the same application, which was seen as inefficient:
 \qqq{I always find it quite strange when they have to ask again.} (P55)\\
 %\qqq{My bank has updated their app three or four times in the last couple of months, and every time I reopen it, you have to agree again and accept their terms and conditions.}(P55)
 %\qqq{You go on a site on a daily basis, you have to agree almost on a daily basis, on the same site, so it's just easy just to kind of just click on the option, just to get through it. Otherwise ... just be wasting so much time.}(P83)
\ \\
\emph{\textbf{(4) Information clarity}} (how easy it is to understand the online consent text)
 Participants felt that online consent text should be understandable also by non-specialist users in technology and data protection.
 \qqq{Someone like me who is not very techy doesn't really understand ... any information.} (P26)
 %\qqq{It should be idiot-proof so that you can read it and go through it and think I've spent 30 seconds here and I've ticked the boxes I've said 'yes' and 'no' and I'm reasonably happy.}(P55)
 \qqq{A bit more simple, a lot less kind of like jargon and lingo}. (P57)\\
\ \\
\emph{\textbf{(5) Summary Information}} (summary consent information, with links to further detail (for example, clickable icons that link to detailed information) [see \textit{Uncertainty Avoidance}]

\subsubsection{Fairness}
\emph{Fairness} was defined as a user's feeling that their personal rewards and costs and those of another party are in balance with each other.  Three value creators contributing to the value of fairness were identified.\\
\ \\
\emph{\textbf{(1) Should Consent be Required?}} (feeling that the user’s need for consent is in balance with the extent of the functionality that the online service or website provides to the user).

 Participants felt that consent procedures should be proportionate int that users' effort should be in balance with the functionality that consent gives access to:
 \qqq{But you know if all I really want to do is look at a picture or read a very brief article, I don't want to read 30 pages of terms and conditions.} (P55)
% \qqq{It would only be if you choose to interact with the services of that website that you then have to give permission for something. So if you want to sign up for an account or you want to use something that they sell or or or whatever, but just to view a website, they don't really need to know.}(P55)
 %\qqq{I often think they do seem to want an awful lot of information just so you can view website.}(P55)
 %\qqq{When you're just a visitor, that - I think - is is something where you shouldn't have to give a lot of information.}(P55)

% In particular for less important or less risky functionality (such as viewing a website) independent reassurance was seen as useful to provide peace of mind to a user if they were not prepared to or capable of reading a long consent document:
 %\qqq{Because a lot of it is, it is very tedious and ... it would be nice if someone had said this has been checked as OK. And then I didn't have to put my faith in that.}(P55)\\
 %\qqq{I think this is where we just kind of accept it and put our faith in it, because at the end of the day, it's a website. It's not very important. So  I suppose I would be less unhappy somebody telling me this is OK than I am now just having to accept it.}(P55)
 \ \\
\emph{\textbf{(2) Functionality}} (feeling that the nature or volume of personal data the user provides is in balance with the functionality they receive from the online service or website)

 Users realised that online services often operate a business where, as a condition for  using an online service for free of monetary change, users give their personal data.  Users consented although they did not (fully) agree:

 \qqq{You make a deal with the devil. You've got to pay the price. So, yeah, I'm kind of one of those where if I want to use a service, I have to accept that I give them my data and they use it the way they see fit.} (P55)
 %\qqq{If you've got to get through something just to read it, I'll just go somewhere else rather than give them all that information.}(P55)
 %\qqq{Feel sometimes as if it's a quid pro quo you give them your data and they give you a service.}(P26)\\
 %\qqq{I'm sure that there are certain parts that I would not be 100\% happy with.  And I think [the] problem is, even if I read through them all, even though there are issues that I am not 100\% happy with, I know that I want the service or purchase what I wanted … then I'm prepared to accept compromise.}(P33)

\emph{\textbf{(3) Legal Protection}} (feeling that the user’s difficulty of understanding the complexity of the text is in balance with protection that the text gives the online service provider against legal action).

 The use of specialist legal language that was challenging for users to understand was seen as unfair:
 \qqq{I think it should be clear enough that you don't have to have done a law course at university to understand the basics.} (P55)
 %\qqq{Sometimes you feel like it's cut and paste.}(P55)
 %\qqq{I say if you do go and read the terms and the conditions, they are lengthy. They start with the first party and the second party and the third party involved. And it may involve a fourth party ... This is like when I did my will or when I got my mortgage and you know I had to read all of those, they were important. I find it quite strange that that amount of complex information is required just to visit a website.}(P55)
 \qqq{It would be important to avoid legalese in the phrasing of these standardised forms.} (P64)
 %\qqq{Sometimes it's quite legalistic as well. So you don't know what you're signing up for as far as that's concerned, there needs to be a lot of work done around that, I think.}(P73)

%Because of specialist legal language readers would not read the text; then given consent could not be meaningful:
% \qqq{If it's just long legal copy I cannot be bothered with that.}(P75)

% Although summarising the relevant legal content of the consent might be beneficial in terms of facilitating users reading and understanding the text, an incomplete summary could cause legal problems:
% \qqq{Could provide a summary of the legal terms-and-conditions text, but this would be legally problematic. This is because potentially not all relevant legal aspects would be covered. Therefore, if a user makes a legal challenge, the service provider would not be fully covered the way they are now with their legally comprehensive (from their perspective) set of terms and conditions.}(P64)

\subsubsection{Loss Aversion}
\emph{Loss aversion} was defined as a  user's preference to avoid losses.  Two value creators contributing to the value of loss aversion were identified.\\
\ \\
\emph{\textbf{(1) Risk Information}} (risk-related data and communication of potential risk)
 A strategy for users to reduce risk was to stay with trusted major companies rather than  rely on consent documents:
 \qqq{I don't even look if I know a company's name ... I tend to just trust it anyway. So that's why I tend to stick with the big names of the brand.} (P57)

% Another strategy was to actively scrutinise a consent document for any risk before making a consent decision, but this was difficult as the process was too time-consuming:\\
% \qqq{Make sure you know what the risks are and then decide whether to take them or not.}(P73)
 %\qqq{I probably would just look at the summary and think what you know, are there risks ? So the the main thing you're looking for is actually potential risks.}(P73)
% \qqq{You've got to tick this box and say that you've read everything, but to be quite frank, there isn't time or to go right through all the fine detail. So I always think it's a bit of a risk, but if you want to go ahead and purchase, you know whatever it is, then really you have to tick it. If you don't want to purchase, then you just have to come out and get it some other way. So I find I find them very difficult. I think that you could go wrong with them because you've ticked a box, but you don't know what you've ticked.}(P73)
\ \\
\emph{\textbf{(2) Information about Consequences}} (information linked to important risky consequences for the user)

 Participants believed that about information risky consequences of consent for online services was not always available to or not read by users, with  serious potential consequences.  In response, a strategy was for users to reduce the personal information they gave away:

 \qqq{I mean you sign up to one site and you didn't know but probably ... access to God knows how many countries ... data being sold to Russia and China or whatever exactly that they would be doing on a daily basis without [you] knowing.} (P77)
 %\qqq{I'm always thought of the  South Park episode where he doesn't read the terms and the conditions and gets taken by Apple. And I think, yeah, it's awful, but really you could do, couldn't you, could sign up to anything without realizing.}(P55) https://en.wikipedia.org/wiki/HumancentiPad
% \qqq{Generally, the main motivation is the easiest option and don't want to give too much away without really kind of knowing I'm giving away.}(P83)

\subsection{STUDY 2 } \label{study2}
Study 2 sought to quantify the perceived relative importance of the values and value creators identified in Study 1.
%hierarchical means-end structure that was  \textcolor{red}{what does quantify mean?}

\emph{Consistency}. Consistency (of comparative judgment) ratios were calculated for the top-level goal of informed online consent from the pairwise compared values and for each of  four of the five values from the pairwise compared values. (Consistency did not apply to the value of loss aversion, as there were two value creators.) According to the standard cut-off for consistency (consistency ratio, CR{\textless} 0.10)  27 pct to 36 pct of cases was consistent, and 63 pct to 69 pct was consistent with a cut-off of 0.20.  

\textit{Weightings, sensitivity analysis}. Sensitivity analysis was conducted in the subsequent analysis of weightings for values and value creators to establish the robustness of the findings.
The pattern of results of means and confidence intervals for informed online consent and each of  the five values was the same for the two cut-offs; the pattern of inferential statistics was the same for the two cut-offs (see results below). 

\begin{figure*}[bp]
\centering
\begin{tabular}{p{0.45\textwidth}|p{0.45\textwidth}}
\hline
   \includegraphics[width=0.4\textwidth]{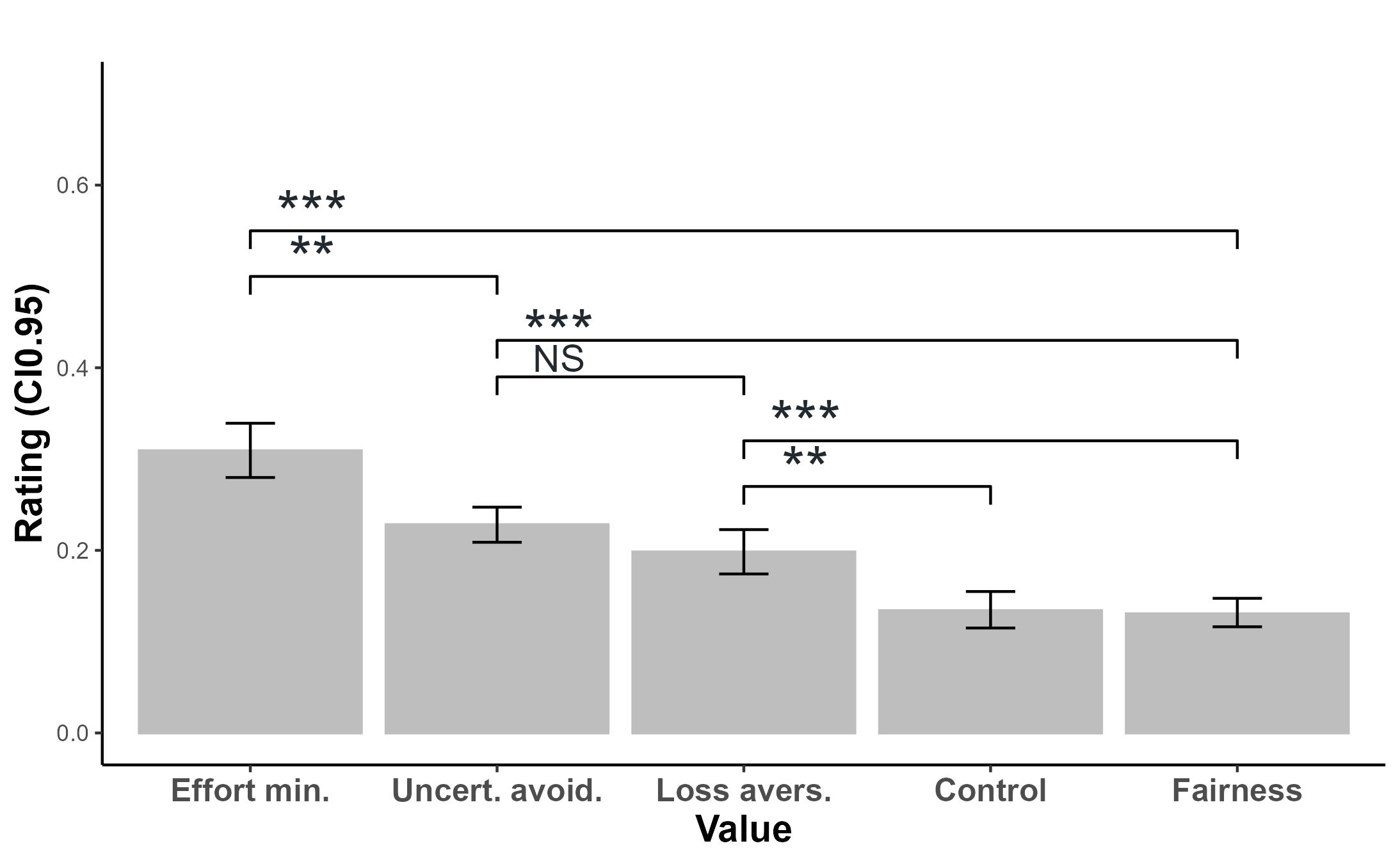}
 {\small{Informed Consent}}

    %\label{fig:FigWeightingsConsent}   %\end{figure}
     & %\begin{figure}
      \includegraphics[width=0.45\textwidth]{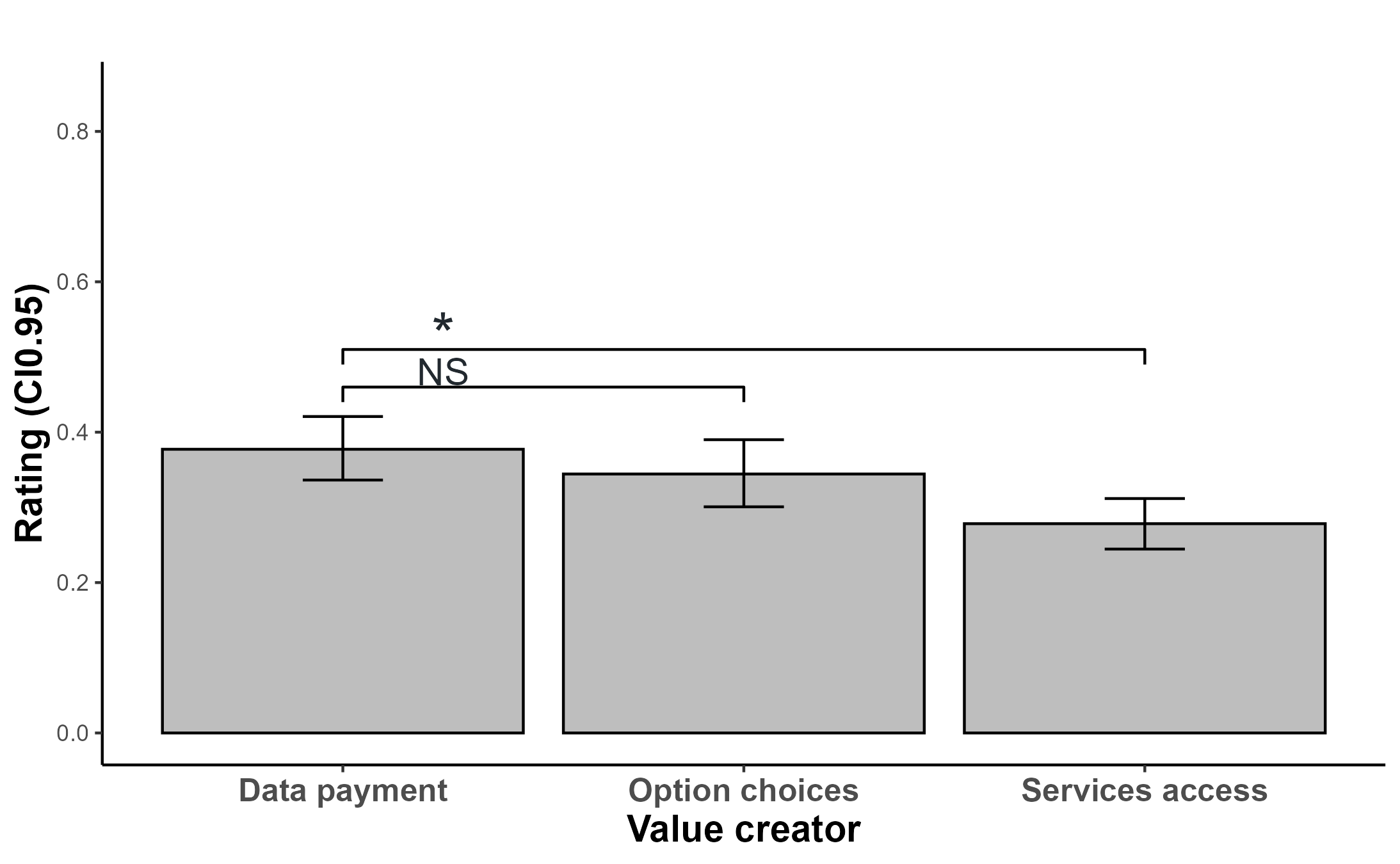}
   {\small{Control}}
 \\ \hline

 \includegraphics[width=0.45\textwidth]{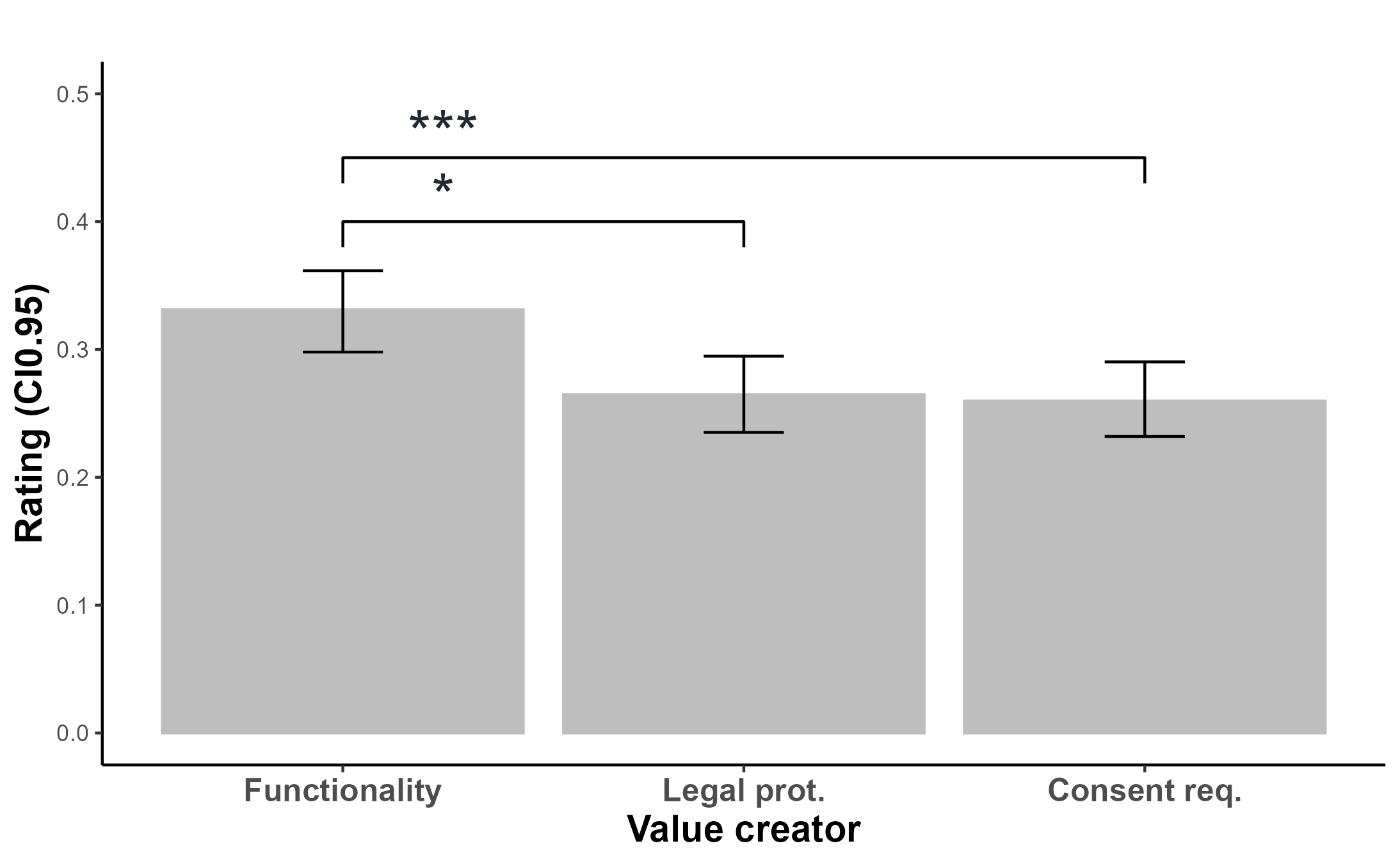}
    {\small{Fairness}}
    &
    
 \includegraphics[width=0.45\textwidth]{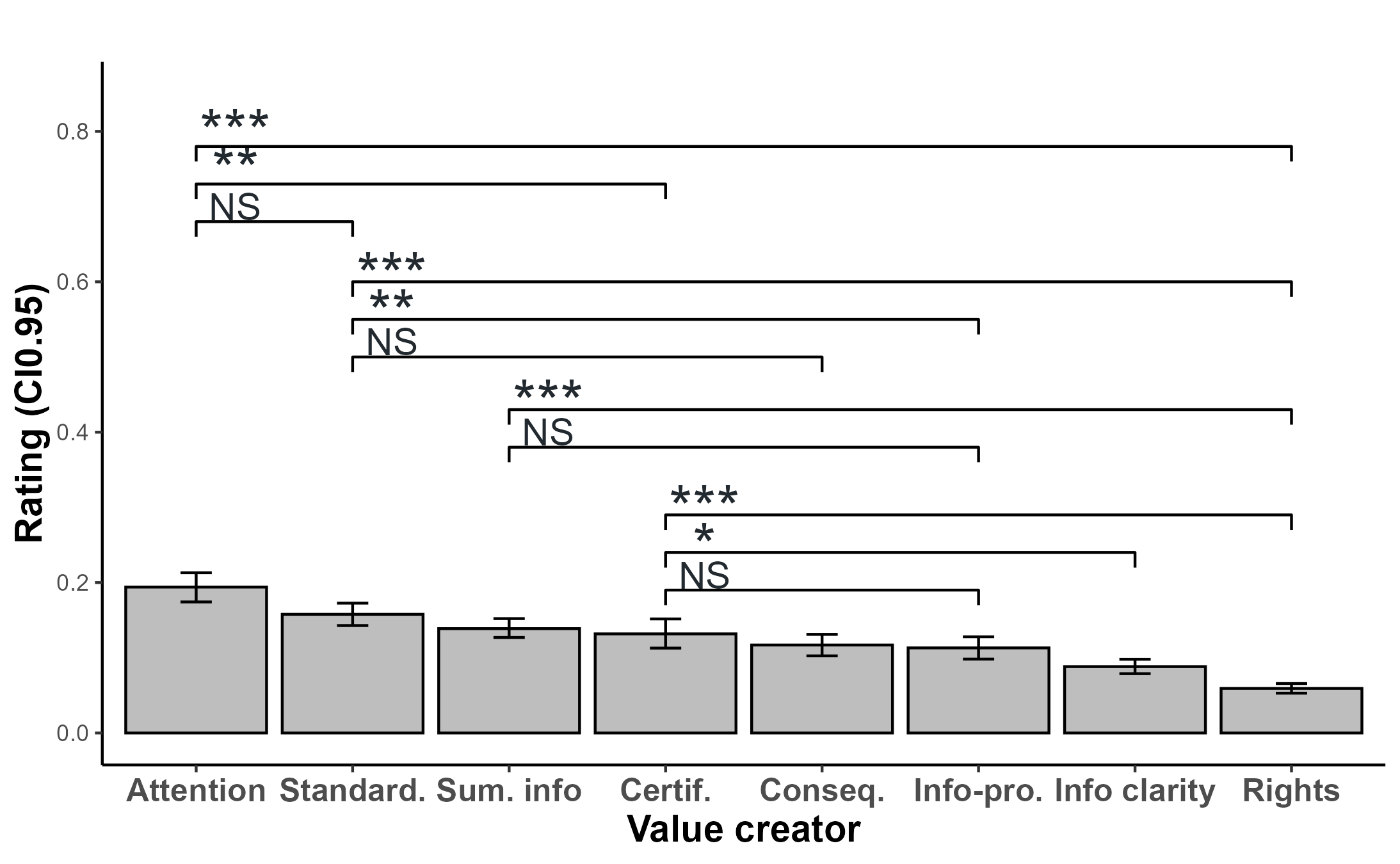}\newline
    {\small{Uncertainty Avoidance 1/2}}\\ \hline

%\includegraphics[width=0.4\textwidth]{figs/weightings.0.10.fairness.png}
%   {Value creator weightings for fairness: means with confidence interval}
%   &
%  \includegraphics[width=0.4\textwidth]{figs/weightings.0.10.uncertainty.avoidance.png}
%    {Value creator weightings for uncertainty avoidance: means with confidence interval}   \\
%\hline

 \includegraphics[width=0.45\textwidth]{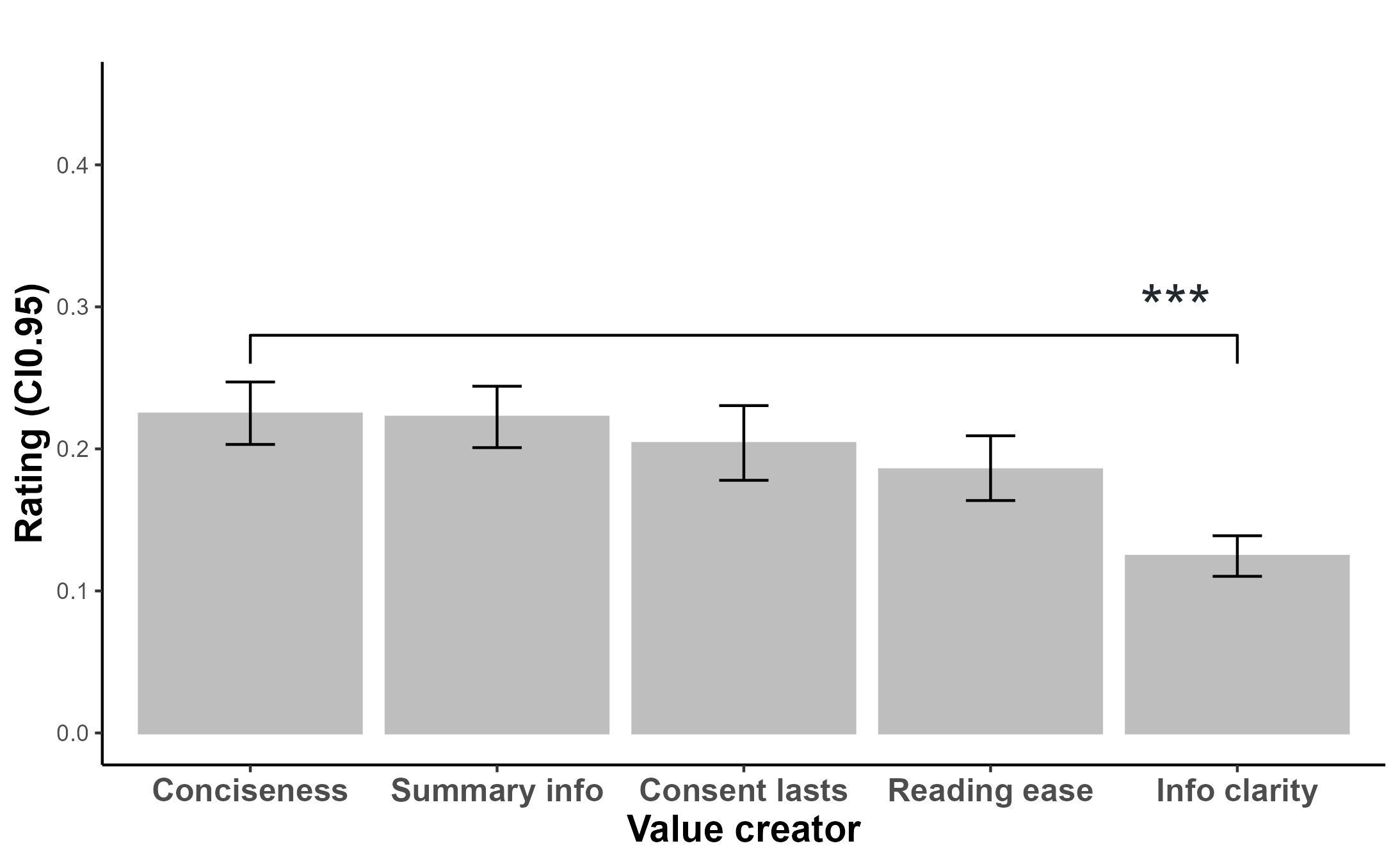}
    {\small{Effort Minimisation}}
    &
  \includegraphics[width=0.45\textwidth]{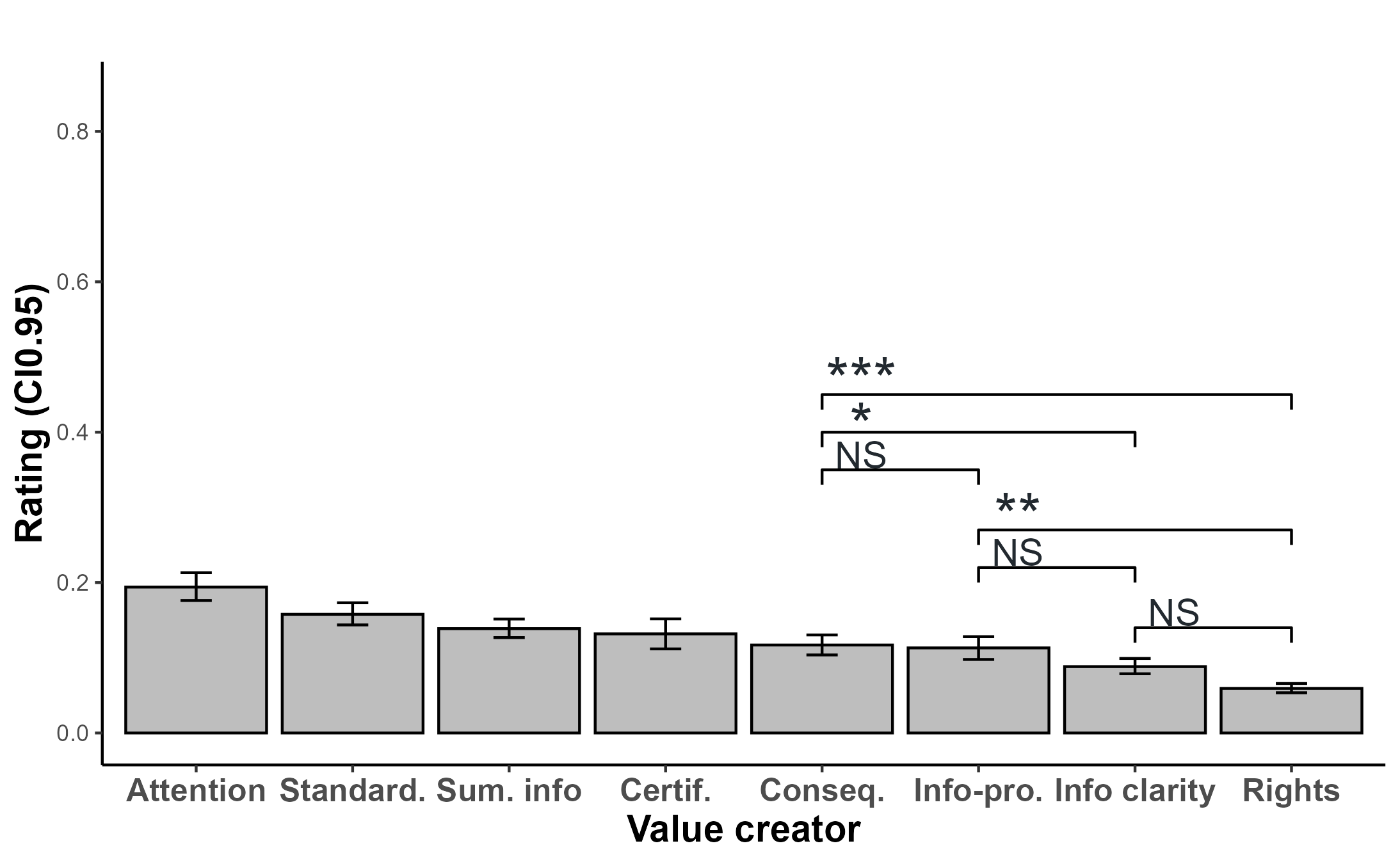}
   {\small{Uncertainty Avoidance 2/2}}\\   \hline
    
\end{tabular}
\caption{Value Creator Weightings for Informed Online Consent Means with Confidence Intervals}
* \textit{p} {\textless} 0.05; ** \textit{p} {\textless} 0.01; *** \textit{p} {\textless} 0.001; NS: not significant.
\label{fig:FigWeightingsConsent}
\end{figure*}

\emph{Weightings}. The means with confidence intervals (Diagrams in Figure \ref{fig:FigWeightingsConsent} %-\ref{fig:FigWeightingsLossAversion}) 
%(Figures \ref{fig:FigWeightingsConsent}, \ref{fig:FigWeightingsControl}, \ref{fig:FigWeightingsFairness}, \ref{fig:FigWeightingsUncertaintyAvoidance}, \ref{fig:FigWeightingsEffortminimisation}, \ref{fig:FigWeightingsLossAversion})
) show substantial variability among the values for informed online consent and among the value creators for each value (except for loss aversion).  
t-tests showed no effect of employment status on any of the weightings. Two-way mixed ANOVA, with Greenhouse-Geisser correction for sphericity, showed \textit{no main effect of employment status and no interaction effect of status with value or value creator on any of the measures}, for informed online consent or any of the four values.  Two-way mixed ANCOVA showed that neither were ATI and OPLIS significant covariates. Therefore, the results of one-way repeated-measures ANOVA with value or value creator as the independent variable, corrected for sphericity, are reported  here. The results for CR {\textless} 0.10 are reported here (the results for CR {\textless} 0.20 [available on request] follow the same pattern).  The main effects of value (for informed online consent) and value creator (for each of the values) are reported here as well as pairwise comparisons.

For \emph{informed online consent}, the effect of value was significant, \emph{F} (3.12, 227.75) = 33.67, \emph{p} {\textless} 0.001, \emph{pes} = 0.32. Effort minimisation was the dominant value (greater mean than that of the other values),  followed by uncertainty avoidance and loss aversion. For the value of\emph{ control}, the effect of value creator was significant, \emph{F} (1.85, 105.62) = 3.74, \emph{p} = 0.03, \emph{pes} = 0.06. The mean for payment for data was greater than that for access to services. For the value of \emph{fairness}, the effect of value creator was significant, \emph{F} (1.97, 140.14) = 8.08, \emph{p} {\textless} 0.001, \emph{pes} = 0.10. Functionality was the dominant value (greater mean than that of the other value creators). For the value of \emph{uncertainty avoidance}, the effect of value creator was significant, \emph{F} (4.89, 246.89) = 20.20, \emph{p} {\textless} 0.001, \emph{pes} = 0.32. Attracting your attention was the most dominant value creator (higher mean than that of the other value creators, except for Standardisation), followed by Standardisation, and summary information and certification. For the value of \emph{effort minimisation}, the effect of value creator was significant, \emph{F} (3.26, 205.53) = 10.29, \emph{p} {\textless} 0.001, \emph{pes} = 0.14. Information clarity had the lowest weighting (mean smaller than that of  the other value creators); the other value creators were not significantly different.  For the value of \emph{loss aversion}, the effect of value creator was not significant, \emph{F} (1, 203) = 2.61, \emph{p} = 0.11, \emph{pes} = 0.01. 

\section{Discussion \& Reflection}% KAREN, with contributions from PAUL}
The relative quantification of values and value creators  shown in Figure \ref{fig:FigWeightingsConsent} is  instructive. In particular, the results show that effort minimisation is most important. This justifies our proposal of reducing length of policies so as to reduce effort. However, the second most important one is uncertainty avoidance, so it is important to ensure that in reducing length we ensure that the information people want is easily accessed. We can use the value creators as a steer in terms of what information people want to see in a consent form.

The surprising finding is related to the relatively low ranking of control. This is interesting because the very consent form mechanism is based on the assumption that users want to control their privacy, in other words have control over who has their data and how these are used \citep{GDPR}. The relatively low ranking of control, accompanied by the low ranking of consumer rights as a value creator, by both unemployed and employed participants, calls this assumption into question. This apparent indifference might be due to the issues mentioned earlier, namely the frequency with which users are presented with these forms, and the effort that is required to process them. It might be that they are making a perfectly reasonable trade-off in order to be able to get anything done at all. 

The other surprising low-ranked value is fairness, because we know that humans have a deep need to be treated fairly \citep{folger1998organizational,folger2001fairness, nicklin2011importance, folger2011social,ganegoda2015framing,folger2019fairness}. Even so, our participants indicated that this value did not mean as much to them as the other values. It might be that people have come to expect to be treated unfairly in this domain, or that effort minimisation and uncertainty avoidance are just that much more important in this context.

%The theoretical psychological basis of control (linked to autonomy = one of the three dimensions of SDT), fairness (Fairness theory; \citep{folger1998organizational, folger2001fairness, nicklin2011importance, folger2011social,ganegoda2015framing,folger2019fairness}, uncertainty avoidance (UA is one of Hofstede's dimensions), effort minimisation (System 1 thinking; scarcity theory), and loss aversion (prospect theory and other)

\section{Conclusion \& Future Work}
We reviewed the research literature on the issues related to consent forms, and suggested a way to improve the situation: a value-driven approach which can be used to remove information (and shorten the policy) by providing information that users value rather than all possible information in the consent form. To apply this approach, we needed to understand what users value. We decided to focus on the values of unemployed users given the particular challenges they face in protecting their privacy online. 

The \emph{raison d'\^{e}tre} for this research was thus to inform a value-driven approach in reducing the length of consent forms, with a specific focus on unemployed users. We carried out two studies which, together, delivered a quantified hierarchy of values and value creators. This will inform the next stage of our study where we will carry out an empirical study to compare user preference and comprehension of a traditional consent form versus a value-driven consent form.

As future work, it would be very interesting and important to explore the reasons behind these rankings so that we understand users' thought processes when contemplating and dealing with online consent forms. 
%\nocite{*}
%\bibliographystyle{SCITEPRESS/apalike}
\bibliographystyle{ACM-Reference-Format}
{\small
\bibliography{refs,refs2}}
%\bibliography{refs,refs2}
%\bibliographystyle{abbrvnat-jpc}

\appendix

\begin{table}[ht]
{\small

    \centering
        \caption{Results of Searches}
    \label{tab:searches}
    \begin{tabular}{|l|c|c|}
    \hline
    \textbf{Database}    &  \textbf{Number} & \textbf{After Exclusion}\\ \hline
    SCOPUS     &14  &9 \\\rowcolor{Gray}
    ProQuest     &1 &0 \\
    EBSCO     & 2&0 \\ \rowcolor{Gray}
    SSRN     & 6& 4\\ 
    Ingenta     & 29& 14\\  \rowcolor{Gray}
    DBLP     & 3&3 \\ 
    SpringerLink     &2 &1 \\  \rowcolor{Gray}
    ACM DL & 86 &11 \\ 
    Google     & 12& 6\\ \hline
    Total & & 48\\ \hline
    \end{tabular}

    }
\end{table}
\end{document}